\documentclass[aps,prl,reprint,showpacs,amsmath,amssymb]{revtex4-1}
\usepackage{graphicx}
\usepackage{xcolor}
\usepackage[colorlinks=true,citecolor=blue,urlcolor=blue,linkcolor=blue]{hyperref}

\definecolor{r}{rgb}{1.0,0.0,0.0}
\definecolor{b}{rgb}{0.0,0.0,1.0}
\definecolor{g}{rgb}{0.0,0.5,0.0}
\definecolor{m}{rgb}{0.75,0.0,0.75}
\definecolor{c}{rgb}{0.0,0.75,0.75}
\definecolor{y}{rgb}{0.75,0.75,0.0}

\newcommand*{\solidrule}[1][5mm]{\rule[0.4ex]{#1}{1.5pt}}
\newcommand*{\dashedrule}{\mbox{\solidrule[1mm]\hspace{1mm}\solidrule[1mm]\hspace{1mm}\solidrule[1mm]}}
\newcommand*{\dottedrule}{\mbox{\solidrule[0.33mm]\hspace{0.5mm}\solidrule[0.33mm]\hspace{0.5mm}\solidrule[0.33mm]\hspace{0.5mm}%
\solidrule[0.33mm]\hspace{0.5mm}\solidrule[0.33mm]\hspace{0.5mm}\solidrule[0.33mm]\hspace{0.5mm}\solidrule[0.33mm]}}
\newcommand*{\dotdashrule}{\mbox{\solidrule[0.83mm]\hspace{0.83mm}\solidrule[0.33mm]\hspace{0.83mm}\solidrule[0.83mm]\hspace{0.83mm}%
\solidrule[0.33mm]\hspace{0.83mm}\solidrule[0.83mm]}}

\newcommand*{\mybullet}{\lower .30ex \hbox{\Large $\bullet$}}
\newcommand*{\mysquare}{\lower .10ex \hbox{\scriptsize $\square$}}
\newcommand*{\mylozenge}{\lower .10ex \hbox{$\lozenge$}}

\begin{document}

\title{Attosecond pulse shaping around a Cooper minimum}

\date{18 September 2013}


\author{S. B. Schoun}
\email{schoun.1@osu.edu}
\author{R. Chirla}
\author{J. Wheeler}
\author{C. Roedig}
\author{P. Agostini}
\author{L. F. DiMauro}
\affiliation{Department of Physics, The Ohio State University, Columbus, OH 43210, USA}

\author{K. J. Schafer}
\author{M. B. Gaarde}
\email{gaarde@phys.lsu.edu}
\affiliation{Department of Physics and Astronomy, Louisiana State University, Baton Rouge, LA 70803, USA}


\begin{abstract}

High harmonic generation (HHG) is used to measure the spectral phase of the recombination dipole matrix element (RDM) in argon over a broad frequency range that includes  the $3p$ Cooper minimum (CM). 
The measured RDM phase  agrees well with predictions based on the scattering phases and amplitudes of the interfering $s$- and $d$-channel contributions to the complementary photoionization process.
The reconstructed  attosecond bursts that underlie the HHG process  show that the derivative of the RDM spectral phase, the group delay, does not have a straight-forward interpretation as an emission time, in contrast to the usual attochirp group delay. 
Instead, the rapid RDM phase variation caused by the CM reshapes the attosecond bursts. 

\end{abstract}

\maketitle

\begin{figure}[b]
\includegraphics[width=\columnwidth]{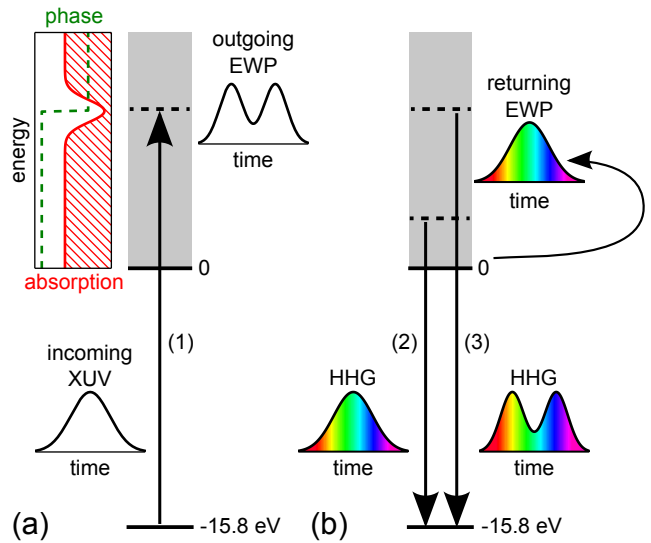}
\caption{\label{cartoon} (Color online) Energy level diagram and time-domain picture of EWPs and broadband XUV pulses in (a) photoionization and (b) photo-recombination in HHG in the vicinity of a Cooper minimum. See text for the discussion of paths (1), (2), and (3). 
}
\end{figure}

One signature of atomic structure, first discussed by Cooper~\cite{cooper1962}, is a local minimum in the photoionization (PI) probability at a specific photon energy.
This Cooper minimum (CM) is caused by a sign change of the bound-free transition dipole in one angular momentum channel and not a modulation of the density of states. 
The CM has been extensively studied using traditional photoionization spectroscopy but the phase of the transition dipole, which undergoes a $\pi$  jump at the CM as illustrated in Fig.~\ref{cartoon}(a),  is not directly accessible, although it strongly influences the measured electron angular distribution and spin polarization~\cite{cherepkov1973, huang1981}. 
The {\it time-domain} consequences of the CM in PI are illustrated  as path (1) in Fig.~\ref{cartoon}(a):  a smooth ultra-fast XUV pulse with sufficient bandwidth and proper photon energy produces an outgoing electron wave packet (EWP) inscribed with the CM hole \cite{hoogenraad1998}. Experimental reconstruction of such an outgoing EWP, however, requires complete knowledge of the amplitude and phase \cite{yakovlev2010}, and to date no measurement has demonstrated the double-peaked structure illustrated in  Fig.~\ref{cartoon}(a).

In high harmonic generation (HHG) the inverse process of broadband PI takes place: an EWP, promoted and accelerated in the continuum by an intense, optical pulse, returns to the core and gives rise to emission of  XUV radiation through photo-recombination, as illustrated by path (2) in Fig.~\ref{cartoon}(b). The EWP thereby acts as a self-probe of the complex  recombination dipole matrix element (RDM). A measurement of the emitted XUV radiation can provide access to both the amplitude and phase of the recombination dipole and can thus be used to study structural features of the generating atom or molecule \cite{itatani2004, mcfarland2008, le2009, haessler2010}, and in particular the CM. Path (3) in  Fig.~\ref{cartoon}(b) illustrates one main point of this paper: that the time domain consequences of the CM amplitude and phase modulation are such that a smooth returning EWP will be shaped into a structured XUV time profile.

In this Letter, we experimentally and theoretically investigate the phase modification in the high harmonic emission induced by the 3$p$ Cooper minimum of argon. 
In the experiment, the derivative of the spectral phase, the group delay (GD), 
is measured using the resolution of attosecond beating by interference of two-photon transitions (RABBITT) method \cite{paul2001,muller2002}. 
The experimental phase and amplitude are compared with both a simple model and a more comprehensive calculation using numerical solutions of the Schr\"{o}dinger  and Maxwell wave equations. 
Previous high harmonic studies \cite{worner2009, farrell2011, higuet2011, jin2011} have established the existence of the CM in the frequency spectrum of the emitted harmonic comb, but not in the spectral phase.
Furthermore, accessing the RDM phase from the RABBITT requires knowledge of the other contributions to the total measured phase. 
The leading contribution is a single atom effect caused by the accumulated phase of the field-driven rescattering electron wave packet \cite{kazamias2004, varju2005}, the so-called attochirp. 
In addition, there are several macroscopic contributions involving the propagation through the argon source, spectral filters, and the RABBITT detection gas. 
The results of this study unequivocally establish that direct measurements of the RDM phase near a CM can be extracted from RABBITT characterization of HHG.

This work also allows us to establish the relation between the measured GD and the time structure of the harmonic dipole emission in the presence of an (anti)resonance.
Typically in HHG, the GD is interpreted as the emission time of different frequencies in the harmonic spectrum, as depicted by path (2) in Fig.~\ref{cartoon}(b). 
This is justified for short trajectory, on-axis emission since the phase originates from the temporal spread of the rescattering EWP in which  higher energy electrons return to the core later than lower energy electrons \cite{mairesse2003, kazamias2004, varju2005}.  
In contrast, the CM phase variation produces an additional structure in the attosecond burst every half cycle of the drive laser field, thus reshaping the temporal envelope of the XUV light. This is depicted as the doublet in path (3) of Fig.~\ref{cartoon}(b).

\begin{figure}[t]
\includegraphics[width=\columnwidth]{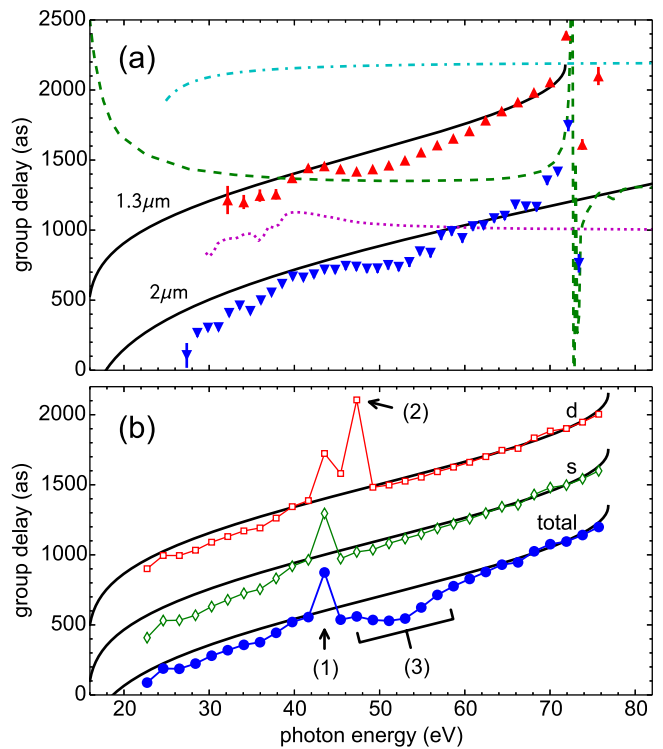}
\caption{\label{GDmeasurement} (Color online) (a) Raw experimental GD  measurements for 1.3~$\mu$m~(\textcolor{r}{$\blacktriangle$}) and 2~$\mu$m~(\textcolor{b}{$\blacktriangledown$}). Estimated GD contributions (offset for clarity) from the Ne atomic phase delay~(\textcolor{c}{\protect\dotdashrule}), dispersion from 0.2~$\mu$m thick Al filter~(\textcolor{g}{\protect\dashedrule}), and neutral Ar dispersion for a gas density-length product of $2.5 \times 10^{17}\, \text{cm}^{-2}$~(\textcolor{m}{\protect\dottedrule}). (b) TDSE-MWE calculations of the total GD~(\textcolor{b}{\protect\mybullet}) for 1.3~$\mu$m, along with the individual  $s$-~(\textcolor{g}{\protect\mylozenge}) and $d$-contributions~(\textcolor{r}{\protect\mysquare}) to the GD.  
In both plots, the attochirp GD~(\protect\solidrule) is shown for reference.
Note that the measured macroscopic GD at 1.3~$\mu$m and 2~$\mu$m differ since they have different attochirp GD and were obtained at different medium lengths (2.5~mm and 0.25~mm respectively), thus  the Ar dispersion is  more apparent in the longer cell 1.3~$\mu$m results. The numbered regions are discussed in the text.
}
\end{figure}

In the experiment, the HHG source utilizes a tunable optical parametric amplifier to generate  65~fs, mid-infrared (MIR) pulses at the wavelengths of 1.3~$\mu$m and 2.0~$\mu$m at a 1~kHz repetition rate.  
The laser beam is focused $\sim\! 1.0$~mm in front of the argon gas jet to phase-match the short trajectory, and, in combination with an aperture in the far-field, minimize long trajectory harmonic contributions.  
A typical argon density in the interaction region of $\text{4--7} \times 10^{18}\, \text{cm}^{-3}$ provides good phase-matching of the highest energy harmonics while reducing reabsorption of the low energy harmonics.

The experimental apparatus for the RABBITT measurement is a Mach-Zehnder interferometer.
The HHG light propagates in one arm and a small amount of the fundamental MIR light propagates in the other.
The pulses are recombined with a variable relative delay, spatially overlapped, and focused into a magnetic bottle electron spectrometer (MBES) \cite{kruit1983}. 
Neon gas introduced into the MBES is photoionized via XUV one-photon and XUV+MIR two-photon processes. 
Information about the XUV spectral phase is encoded in the delay dependent oscillation of the sidebands arising at even harmonic orders in the photoelectron energy spectrum. We note that the use of the longer wavelength MIR sources in this study provides several advantages over conventional 0.8~$\mu$m drive lasers for characterizing the CM amplitude and phase. 
First, the harmonic cutoff energy scales quadratically with the drive laser wavelength \cite{tate2007}, thus allowing an easy extension of the harmonic spectral plateau well beyond the argon CM structure at 50~eV photon energy. 
Consequently, in the region of interest the attochirp GD is well approximated as  linear \cite{doumy2009}. 
Second, the smaller photon energy allows for a finer sampling of the CM phase in the frequency domain.

\begin{figure}[t]
\includegraphics[width=\columnwidth]{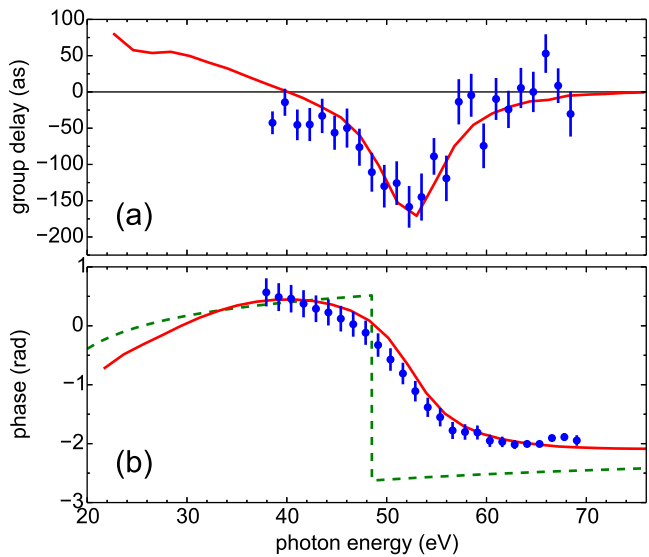}
\caption{\label{dipoleGD} (Color online) Argon measurement~(\textcolor{b}{\protect\mybullet}) and TDSE-MWE calculation~(\textcolor{r}{\protect\solidrule}) of the RDM (a) group delay and (b) phase. 
(b) also plots the scattering-phase difference $\xi$~(\textcolor{g}{\protect\dashedrule}) used in the simple model, Eq.~(\protect\ref{model_eq}), where the instantaneous $\pi$ jump has been added in by hand.
}
\end{figure}

Figure~\ref{GDmeasurement}(a) presents the measured GD as a function of photon energy for two different driving wavelengths, 1.3~$\mu$m and 2~$\mu$m. 
The peak laser intensity in the interaction region is estimated to be 110~TW/cm$^{2}$ and 94~TW/cm$^{2}$, respectively.  
Also plotted are the estimated contributions 
to the total measured GD that are used to retrieve the RDM contribution:  the attochirp GD of the short trajectory calculated semi-classically as in  \cite{kazamias2004, varju2005}, the macroscopic GD including neutral-atom \cite{kim2007} and  aluminum filter dispersion \cite{lopez-martens2005}, and the atomic phase delay \cite{veniard1996,toma2002, mauritsson2005, dahlstrom2013} of the RABBITT detection gas \footnote{One additional contribution to the measured GD which has not been estimated is the effect of the energy dependence of the harmonic Gouy phases as recently measured in \protect\cite{frumker2012}. We speculate that this may contribute to the poor experimental agreement with the classical wave packet group delay for harmonics below 38~eV.}.  
After subtracting these various contributions from the 2~$\mu$m total measured GD, we arrive at the main result of this paper: the RDM group delay shown in Fig.~\ref{dipoleGD}(a) and, upon integration, the RDM phase in Fig.~\ref{dipoleGD}(b) near the CM antiresonance.

The large-scale calculations in this study are performed by solving the coupled time-dependent Schr\"{o}dinger equation (TDSE) and  Maxwell's wave equation (MWE). The MWE is solved by space-marching through the harmonic gas jet with each plane in the propagation direction having a source polarization field term calculated by numerically solving the TDSE  \cite{farrell2011, gaarde2011} in the single active electron approximation. We use an $\ell$-dependent pseudo-potential to describe the electron-core interaction. 
 This  allows us to  separately calculate the major contributions to the dipole moment (and thereby the macroscopic electric field) which come from transitions from continuum $s$- and $d$-states to the ground $3p$ state \cite{farrell2011}.  
Tabulated values are used for the linear absorption and dispersion in argon \cite{henke1993}, as these are underestimated by the  pseudo-potential. The GD is calculated from the radially averaged spectral phase of the macroscopic electric field, which is spatially filtered in the far field, like in the experiment.

The TDSE-MWE  curves in Fig.~\ref{GDmeasurement}(b) illustrate the origin of the CM group delay and the  agreement with the measurement.
Similar to the experiment, the GD is calculated assuming 130~TW/cm$^2$, 1.3~$\mu$m driving field focused 1~mm before the center of a 1.25~mm Ar gas jet with a density of $2.5 \times 10^{18}$~cm$^{-3}$. 
The agreement of the measured and calculated total GD  is excellent except near  $\sim$~44~eV photon energy where the theory overestimates the dispersion contribution  \footnote{In the region where  linear dispersion is large, the calculated spectrum often has multiple peaks. This leads to an unstable phase and is the origin of the jump at 44~eV. This also means that irrespective of the  length of the Ar medium in the calculation, the effective length is less than 1~mm}. 
Figure \ref{GDmeasurement}(b) also shows the individual $s$- and $d$-contributions to the GD. 
The $s$-channel shows a simple monotonic increase except for the dispersion jump near 44~eV (indicated by arrow 1).  The $d$-channel GD is similar, with the addition of one large jump for the sideband at 48~eV (arrow 2), owing to the derivative of the instantaneous $\pi$ phase jump in that channel. 
However, the total GD is the coherent sum of both channels, which reproduces the measured CM phase over the range of 48--60~eV (region 3).
Note that the CM influence on the total GD occurs at higher energy than the pure $d$-channel phase jump. A similar effect has been observed in the location of the spectral amplitude minimum of argon HHG \cite{worner2009, farrell2011, higuet2011, jin2011}. 
Since the two channels have the same attochirp, the CM dipole GD can be extracted from the calculation by subtracting the macroscopic $s$-GD from the macroscopic total-GD  \footnote{This works because the dipole GD from the $s$-channel alone is nearly constant (see text surrounding Eq.~(\protect\ref{model_eq})). Had it not been constant it would have to be added back in to obtain the total dipole GD.}. 
The result is shown as the solid lines in Fig.~\ref{dipoleGD} and is in excellent agreement with the experimental RDM which was extracted in a completely different manner. 
This provides clear evidence that RDM phases can be accurately measured using HHG.

\begin{figure}[t]
\includegraphics[width=\columnwidth]{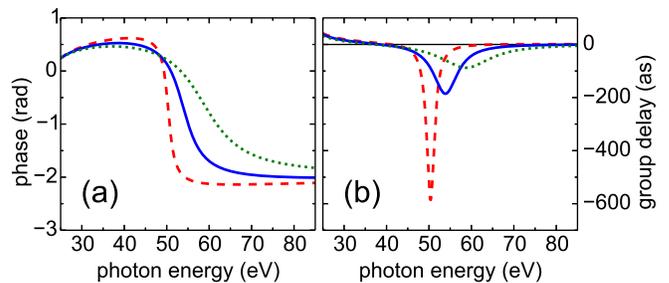}
\caption{\label{SimpleModel} (Color online) Dipole phase (a) and GD (b) calculated from the simple model in Eq.~(\protect\ref{model_eq}) when using $\Delta\omega=2.2$~eV~(\textcolor{r}{\protect\dashedrule}), 6.6~eV~(\textcolor{b}{\protect\solidrule}), and 13.2~eV~(\textcolor{g}{\protect\dottedrule}).
}
\end{figure}

As seen in Fig.~\ref{dipoleGD}(b), the phase transition of the total RDM is  spread over a 20~eV spectral range, and equals about 2.6 radians, even though the $d$-channel RDM undergoes a sharp $\pi$ phase shift at the CM. 
This is consistent with previous calculations  \cite{jin2011, worner2009, kheifets2013, mcfarland2009}. We can use a simple model to provide some additional insight into this phase behavior. 
The total dipole moment $p(\omega)$ is approximated by accounting for the relative strength, $\chi(\omega)$, and the relative scattering phase, $\xi(\omega)$, of the $s$- and $d$-contributions:
\begin{eqnarray}
\label{model_eq}
p(\omega) &=& s(\omega){\rm e}^{i\eta_{0}(\omega)} \left[1+\chi(\omega){\rm e}^{i\xi(\omega)}\right]  \\
\chi(\omega) &=& \frac{d(\omega)}{s(\omega)}, 
~~\xi(\omega) = \eta_2(\omega)-\eta_0(\omega). \nonumber
\end{eqnarray}
We model the relative amplitude as a linear function, $\chi(\omega) = -(\omega-\omega_C)/\Delta\omega$, where
$\omega_C = 48.5$~eV is the energy of the $d$-channel zero crossing and $1/\Delta\omega$ defines the slope, which is estimated to be $\Delta\omega = 6.6$~eV from the macroscopic calculations. 
The  scattering phases  $\eta_{\ell}(\omega)=\sigma_{\ell}(\omega)+\delta_{\ell}(\omega)$ include the Coulomb contribution $\sigma_{\ell}(\omega) = \arg[\Gamma(\ell+1-i/k(\omega))]$, where $k(\omega)$ is the electron wave number in atomic units and $\Gamma$ is the gamma-function, and the short range potential scattering phases ($\delta_{2}=1.4$~rad, $\delta_{0}=0$~rad) which are assumed constant. 
$\xi(\omega)$ is plotted as a dashed line in Fig.~\ref{dipoleGD}(b) with the $\pi$ phase jump due to the sign change of $\chi(\omega)$ also included.

\begin{figure}[t]
\includegraphics[width=\columnwidth]{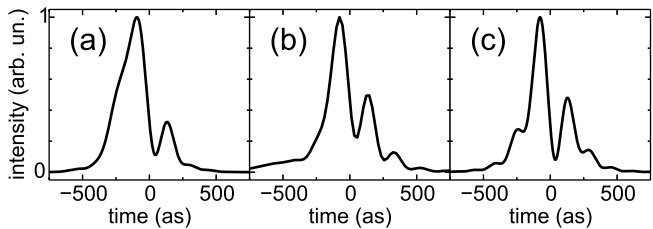}
\caption{\label{timeprofiles} Attosecond XUV time profiles reconstructed from a 24 eV spectral bandwidth centered on the CM feature, using the  amplitude and phase results of the (a) 2~$\mu$m experiment, (b) TDSE-MWE calculations, and (c) simple model, Eq.~(\protect\ref{model_eq}), including the classical wave packet phase to match the attochirp of the experiment. 
}
\end{figure}

The simple model emphasizes that the phase of the total dipole moment is heavily influenced by the relative strength of the two channels, and that the $s$-contribution moderates and shapes the effect of the $d$ phase jump. 
Figure~\ref{SimpleModel} illustrates the sensitivity of the dipole phase and corresponding GD to the relative strength, through the size of $\Delta\omega$.  For example, the (negative) GD  would become arbitrarily large and narrow in Fig.~\ref{SimpleModel} if the $\sim\! \pi$ phase transition were more instantaneous (smaller $\Delta\omega$, corresponding to a weaker $s$-contribution).

Finally, we comment on the time structure of the emitted XUV radiation and, in particular, how the presence of a CM in the dipole moment can alter the shape of the emitted attosecond pulses. 
 Figure~\ref{timeprofiles}(a) shows the time profile of one attosecond burst reconstructed from the 2~$\mu$m experimental spectral amplitude and phase.
The measurements are corrected for the Al filter transmission and dispersion, and the RABBITT detection gas. 
The corresponding  profile  from the macroscopic electric field calculated by the TDSE-MWE is shown in Fig.~\ref{timeprofiles}(b) and yields excellent agreement with the experiment. Also the simple model discussed above, when combined with the attochirp, exhibits a similar, double-peaked time profile as shown in Fig.~\ref{timeprofiles}(c).

The structured attosecond emission shown in Fig.~\ref{timeprofiles} is characteristic of the amplitude and phase modulation caused by a Cooper minimum or, more generally, an antiresonance. 
The beating between the frequency components on either side of the spectral minimum, which are  out of phase by approximately $\pi$, produces multiple peaks in the time domain. 
Thus, the CM shapes the time-dependent dipole and imparts an additional amplitude modulation in the emission not found in the rescattering EWP. 
The same principle, but in reverse, holds for  photoionization, as illustrated in Fig.~\ref{cartoon}(a).
Broadband XUV light tuned near the CM would create an EWP that has additional structure not found in the excitation pulse \cite{hoogenraad1998}. 
In this case one could expect that defining an ionization delay time via the calculated group delay, as is often done, will lead to difficulties, especially if the group delay is larger than the pulse width \cite{serov2013}. 
In contrast, a PI experiment performed with narrow-band XUV light would allow for a much closer correspondence between measured group delay and ionization delay times \cite{haessler2011, caillat2011}.

In summary,  the spectral phase of the RDM in argon around the 3$p$ Cooper minimum has been characterized using RABBITT measurements of the high harmonic emission.
The results show that the characteristic phase structure imposed by the Cooper minimum on the RDM can be extracted from the experimentally measured group delay. 
The measured spectral amplitudes and phases allowed us to reconstruct the time-dependent dipole emission of the attosecond bursts produced every half-MIR-cycle during the harmonic generation process.

\begin{acknowledgments}
The work at OSU was supported by the DOE under contract DE-FG02-04ER15614. The work at LSU was supported by the NSF under Grants Nos. PHY-0701372 and PHY-1019071. High-performance computational resources were provided by Louisiana Optical Network Initiative (www.loni.org). LFD acknowledges support from the Dr. Edward and Sylvia Hagenlocker chair.
\end{acknowledgments}

\bibliography{References}

\end{document}